\documentstyle[12pt]{article}
\input epsf.sty

\catcode`@=11
\def\chkspace{%
  \relax   
  \begingroup\ifhmode\aftergroup\dochksp@ce\fi\endgroup}
\def\dochksp@ce{%
  \unskip              
  \futurelet\chkspct@k\d@chkspc  
}
\def\d@chkspc{%
  \let\nxtsp@ce=\relax
  \ifx\chkspct@k.\else     
    \ifx\chkspct@k,\else
      \ifx\chkspct@k;\else
        \ifx\chkspct@k!\else
          \ifx\chkspct@k?\else
            \ifx\chkspct@k:\else
              \ifx\chkspct@k)\else
              \ifx\chkspct@k(\else
                \ifx\chkspct@k]\else
                  \ifx\chkspct@k-\else
                    \ifx\chkspct@k\egroup\else  
                      \let\nxtsp@ce=\put@space  
                    \fi
                  \fi
                \fi
              \fi
              \fi
            \fi
          \fi
        \fi
      \fi
    \fi
  \fi
  \nxtsp@ce
}
\def\put@space{$\;$}
\catcode`@=12

\def\Pade{Pad$\acute{\rm e}$\chkspace}

\def\ra{{$\rightarrow$}\chkspace}
\def\etal{{\it et al.}\chkspace}

\def\eg{{\it eg.}\chkspace}

\def\apriori{{\it a priori}\chkspace}

\def\ep{{e$^+$e$^-$}\chkspace}
\def\epa{{e$^+$e$^-$ annihilation}\chkspace}

\def\qu{\quad}

\def\gluino{\relax\ifmmode \tilde{g} \else $\tilde{g}$ \fi\chkspace}

\def\qq{q$\overline{\rm q}$\chkspace}

\def\bb{\relax\ifmmode {\rm b}\bar{\rm b}
       \else ${\rm b}\bar{\rm b}$ \fi\chkspace}
\def\cc{\relax\ifmmode {\rm c}\bar{\rm c}
       \else ${\rm c}\bar{\rm c}$ \fi\chkspace}
\def\tt{\relax\ifmmode {\rm t}\bar{\rm t}
       \else ${\rm t}\bar{\rm t}$ \fi\chkspace}
\def\ss{\relax\ifmmode {\rm s}\bar{\rm s}
       \else ${\rm s}\bar{\rm s}$ \fi\chkspace}

\def\qqg{\relax\ifmmode {\rm q}\overline{\rm q}{\rm g}
\else q$\overline{\rm q}$g \fi\chkspace}
\def\bbg{{b$\overline{\rm b}$g}\chkspace}

\def\afb{\relax\ifmmode A_{FB} \else
{{$A_{FB}$}}\fi\chkspace}
\def\afbb{\relax\ifmmode A_{FB}^b \else
{{$A_{FB}^b$}}\fi\chkspace}
\def\pafb{\relax\ifmmode \tilde{A}_{FB} \else
{{$\tilde{A}_{FB}$}}\fi\chkspace}
\def\pafbb{\relax\ifmmode \tilde{A}_{FB}^b \else
{{$\tilde{A}_{FB}^b$}}\fi\chkspace}

\def\pafbzo{\relax\ifmmode \tilde{A}_{FB}|_{O(0)} \else
{{$\tilde{A}_{FB}|_{O(0)}$}}\fi\chkspace}
\def\pafbfo{\relax\ifmmode \tilde{A}_{FB}|_{\oalp} \else
{{$\tilde{A}_{FB}|_{\oalp}$}}\fi\chkspace}
\def\pafbso{\relax\ifmmode \tilde{A}_{FB}|_{\oalpsq} \else
{{$\tilde{A}_{FB}|_{\oalpsq}$}}\fi\chkspace}
\def\pafbto{\relax\ifmmode \tilde{A}_{FB}|_{\oalpc} \else
{{$\tilde{A}_{FB}|_{\oalpc}$}}\fi\chkspace}

\def\pafbbzo{\relax\ifmmode \tilde{A}_{FB}^b|_{O(0)} \else
{{$\tilde{A}_{FB}^b|_{O(0)}$}}\fi\chkspace}
\def\pafbbfo{\relax\ifmmode \tilde{A}_{FB}^b|_{\oalp} \else
{{$\tilde{A}_{FB}^b|_{\oalp}$}}\fi\chkspace}
\def\pafbbso{\relax\ifmmode \tilde{A}_{FB}^b|_{\oalpsq} \else
{{$\tilde{A}_{FB}^b|_{\oalpsq}$}}\fi\chkspace}
\def\pafbbto{\relax\ifmmode \tilde{A}_{FB}^b|_{\oalpc} \else
{{$\tilde{A}_{FB}^b|_{\oalpc}$}}\fi\chkspace}

\def\afbo0{\tilde{A}_{FB}|_{O(0)}}
\def\afbo1{\tilde{A}_{FB}|_{\oalp}}
\def\afbo2{\tilde{A}_{FB}|_{\oalpsq}}
\def\afbo3{\tilde{A}_{FB}|_{\oalpc}}

\def\lam{\relax\ifmmode \Lambda_{\overline{MS}}
       \else {{$\Lambda_{\overline{MS}}$}}\fi\chkspace}
\def\lamuds{\relax\ifmmode \Lambda^{(3)}_{\overline{MS}}
       \else {{$\Lambda^{(3)}_{\overline{MS}}$}}\fi\chkspace}
\def\lamudsc{\relax\ifmmode \Lambda^{(4)}_{\overline{MS}}
       \else $\Lambda^{(4)}_{\overline{MS}}$\fi\chkspace}
\def\lamudscb{\relax\ifmmode \Lambda^{(5)}_{\overline{MS}}
       \else $\Lambda^{(5)}_{\overline{MS}}$\fi\chkspace}

\def\alp{\relax\ifmmode \alpha_s\else $\alpha_s$\fi\chkspace}
\def\alpbar{\relax\ifmmode \overline{\alpha_s}
       \else $\overline{\alpha_s}$\fi\chkspace}
\def\alpmz{\relax\ifmmode \alpha_s(M_Z)\else $\alpha_s(M_Z)$\fi\chkspace}
\def\alpmzsq{\relax\ifmmode \alpha_s(M_Z^2)
       \else $\alpha_s(M_Z^2)$\fi\chkspace}

\def\oalp{\relax\ifmmode O(\alpha_s)\else{{O($\alpha_s$)}}\fi\chkspace}
\def\oalpsq{\relax\ifmmode O(\alpha_s^2)
           \else{{O($\alpha_s^2$)}}\fi\chkspace}
\def\oalpc{\relax\ifmmode O(\alpha_s^3)
           \else{{O($\alpha_s^3$)}}\fi\chkspace}
\def\oalpf{\relax\ifmmode O(\alpha_s^4)
           \else{{O($\alpha_s^4$)}}\fi\chkspace}

\def\plb{Phys. Lett.\chkspace}
\def\npb{Nucl. Phys.\chkspace}
\def\rmp{Rev. Mod. Phys.\chkspace}
\def\prl{Phys. Rev. Lett.\chkspace}
\def\prd{Phys. Rev.\chkspace}
\def\zpc{Z. Phys.\chkspace}

\def\z0{{$Z^0$}\chkspace}
\def\Dst{\relax\ifmmode {\rm D}^* \else {D$^*$}\fi\chkspace}
\def\Dpl{\relax\ifmmode {\rm D}^+ \else {D$^+$}\fi\chkspace}
\def\D0{\relax\ifmmode {\rm D}^0 \else {D$^0$}\fi\chkspace}
\def\Kst{\relax\ifmmode {\rm K}^* \else {K$^*$}\fi\chkspace}
\def\K0{\relax\ifmmode {\rm K}^0_s \else {K$^0_s$}\fi\chkspace}
\def\Kpl{\relax\ifmmode {\rm K}^+ \else {K$^+$}\fi\chkspace}
\def\Kstz{\relax\ifmmode {\rm K}^{*0} \else {K$^{*0}$}\fi\chkspace}

 1

\catcode`\@=11 
\makeatletter
\def\@seccntformat#1{\csname the#1\endcsname.\hskip 1em}

\makeatother
\begin{document}

\hfill{SLAC--PUB--7631}
 
\hfill{MIT-LNS-97-272}
 
\vskip .1truecm

\hfill{September 1997}
 
\vskip 1truecm

\centerline{\Large \bf RESULTS ON \alp AND QCD}
\centerline{\Large \bf FROM (AND ABOVE) THE \z0$^{\star}$}
  
\vskip 1.5truecm
 
\centerline{\bf P.N. Burrows$^{**}$}
 
\vskip .5truecm
 
\centerline{\it Stanford Linear Accelerator Center}
\centerline{\it Stanford University, Stanford, CA94309, USA}
 
\vskip .4truecm
 
\centerline{burrows@slac.stanford.edu}
  
\vskip 1.4truecm
  
\centerline{ABSTRACT}

\noindent
Measurements of \alp from \epa experiments are reviewed and compared
with measurements from other processes. Highlights 
are presented of recent QCD studies in \epa at the \z0 resonance.

\vskip 1truecm

\noindent
\centerline{\it Invited talk at}
\centerline{\it XVII International Conference on Physics in Collision,} 
\centerline{\it Bristol, England, June 25-27, 1997}
 
\vfill
 
\noindent
$\star$ {Work supported by Department of Energy contracts
DE--FC02--94ER40818 (MIT)
and   DE--AC03--76SF00515 (SLAC).}
 
\noindent
$**$ {Permanent address: Lab. for Nuclear Science,
M.I.T., Cambridge, MA 02139, USA.}
 
\eject
  
\section{Introduction}

In electron-positron annihilation 
hadronic activity is, by construction, limited to the final state, making
the study of hadronic events cleaner and simpler relative to lepton-hadron
and hadron-hadron collisions, from both the experimental and
theoretical points-of-view. On the experimental side there are no remnants of
the beam particles to add confusion to the interpretation of hadronic
structures, and, apart from initial and final-state photon radiation effects,
the hadronic centre-of-mass frame coincides with the laboratory frame.
On the theoretical side the absence of hadrons in the incoming beams
removes dependence on the limited knowledge of the parton density 
functions of hadrons, as well as rendering QCD calculations at a
given order of perturbation theory easier to perform because there are
generally fewer strong-interaction Feynman diagrams to consider.

To be specific, samples of hadronic events can be selected by experiments 
at the \z0 resonance with efficiency and purity of better than 99\%. Jet
and event-shape observables have been calculated at next-to-leading order,
\oalpsq, and some inclusive observables have been calculated at \oalpc.
Non-perturbative calculations, in the form of `power corrections' to
perturbatively-evaluated observables, have been performed, and there are
well-understood models of hadronisation that have been carefully tuned to
the data collected over the past 20 years from experiments at the SPEAR, 
CESR/DORIS, PETRA/PEP, TRISTAN, and SLC/LEP colliders.  
Electron-positron annihilation thus provides an ideal environment for precise 
tests of QCD, and has yielded spectacular results which include the first
observation of jets at SPEAR, the direct observation of the gluon at PETRA,
and the detailed study of jets of different flavour at SLC and LEP.

It is impossible to review this wealth of information in the time allotted
here; for a more pedagogical review see Ref.~\cite{ssi}. Instead I have chosen
some `highlights of the past decade' that I feel have significantly advanced
our understanding of strong-interaction physics in the years since
PETRA and PEP. Ten years ago the world's highest-energy \ep data 
sample comprised a few hundred thousand events clustered around a c.m.
energy $Q$ $\sim$ 30 GeV. Since then roughly 16 million hadronic events
have been collected at the \z0 resonance, and there are, in addition, recent
data taken at energies as high as 172 GeV. Furthermore, whereas 10 years ago 
most QCD studies were `flavour blind', the advent of precise silicon-based
vertex detectors has made the separation of light (u,d,s), c and b events,
with high efficiency and purity, relatively straightforward today. 

I have arbitrarily divided my selection of highlights into the 3-jet and
4-jet sectors. The former 
includes precise measurements of \alp and study of
the running of \alp, tests of the flavour-independence of strong interactions,
measurement of quark- and gluon-jet differences, and the search for $T_N$-odd
effects. The latter category includes the 4-jet cross-section and
angular correlations among the jets, which are relevant for searches for
possible beyond-Standard Model (SM) particles known as light gluinos.
 
\section{Measurements of \alp in \ep Annihilation}

\subsection{Theoretical Considerations}
 
The theory of strong interactions, Quantum Chromodynamics (QCD), contains 
in principle only one free parameter, the
strong coupling \alp. QCD can hence be tested in a quantitative fashion
by measuring \alp in different processes and at different hard scales $Q$. 
In practice most QCD calculations of
observables are performed using finite-order perturbation theory, and
calculations beyond leading order depend on the
renormalisation scheme employed. It is conventional to work in
the modified minimal subtraction scheme ($\overline{MS}$
scheme) \cite{msbar}, and to use the strong interaction scale
\lam for five active quark flavours.
If one knows \lam one may calculate the strong coupling \alp($Q^2$) from
the solution of the QCD renormalisation group equation \cite{ian}:

\begin{eqnarray}
\alp(Q^2)\quad=\quad{4\pi\over{\beta_0{\rm ln}(Q^2/\lam^2)}}
\lbrace \;1\;-\;{2\beta_1\over \beta_0^2}\;{{\rm ln(ln}(Q^2/\lam^2))
\over {\rm ln} (Q^2/\lam^2)}\;+\ldots\;\rbrace
\end{eqnarray}

\noindent
Because of the large data samples taken in \epa at the \z0 resonance,
it has become conventional to use as a yardstick \alpmzsq,
where $M_Z$ is the mass of the \z0 boson; $M_Z$ $\approx$ 91.2 GeV.
Tests of QCD can therefore be quantified in terms of the consistency
of the values of \alpmzsq measured in different experiments. 
Measurements of \alp have been performed in \epa, hadron-hadron
collisions, and deep-inelastic lepton-hadron scattering,
covering a range of $Q^2$ from roughly 1 to $10^5$ GeV$^2$; for a recent review
see Ref.~\cite{cracow}.
Some excitement in this area has been generated in recent years due to claims
of an `\alp crisis', see \eg Ref.~\cite{shifman}, with potential implications 
for beyond-Standard-Model physics.

In \epa \alpmzsq has been measured from inclusive observables relating to the
\z0 lineshape and to hadronic decays of the $\tau$ lepton, as well as from
jet-related hadronic event shape observables, and scaling violations in
inclusive hadron fragmentation functions.
 
\subsection{$R$ and the \z0 Lineshape}

For the inclusive ratio $R$ =
$\sigma$(\ep \ra hadrons)/$\sigma$(\ep \ra $\mu^+\mu^-$), the SM electroweak
contributions are well understood theoretically and the perturbative
QCD series has been calculated
up to \oalpc \cite{thirdo} for massless quarks, and up to \oalpsq
including quark mass effects \cite{kuhn}. 
Closely-related observables at the \z0 resonance are:
the \z0 total width, $\Gamma_Z$,
the pole cross section, $\sigma_h^0\equiv
12\pi\Gamma_{ee}\Gamma_{had}/{M_Z^2\Gamma_Z^2}$, and
the ratio of hadronic to leptonic \z0 decay widths
$R_l\;\equiv\;\Gamma_{had}/\Gamma_{ll}$. These are all related to the
\z0 hadronic width:

\begin{eqnarray}
\Gamma_{had}\qu=\qu1.671\left(1\;+\;a_1
\left({\alp\over\pi}\right)
\;+\;a_2\left({\alp\over\pi}\right)^2
\;+\;a_3\left({\alp\over\pi}\right)^3\;+\qu\ldots\right)
\end{eqnarray}

\noindent
where: $a_1$ = 1, $a_2$ = 0.75 and $a_3$ = $-15.3$. 
 
The procedure adopted \cite{blondel} is to perform a global SM fit
to a panoply of electroweak data that includes the W boson and top quark masses
as well as the \z0 lineshape, left-right production asymmetry,
branching ratios to heavy quarks, 
forward-backward asymmetries of final-state fermions,
and polarisation of final-state $\tau$s. The free parameters are
the Higgs mass, $M_{Higgs}$, and \alpmzsq. 
Data presented at the 1996 summer conferences yield
the results shown in Fig.~1~\cite{blondel}, updated for the 1997 Winter 
conferences to yield~\cite{mor97}
the positively-correlated results $M_{Higgs}=127^{+127}_{-72}$ GeV and
\begin{eqnarray}
\alpmzsq\qu=\qu0.120\pm0.003\;\; ({\rm exp.})\pm0.002\;\; ({\rm theor.})
\end{eqnarray}
The \alpmzsq value is lower than the corresponding results
presented at the 1995 conferences~\cite{renton}, \alpmzsq =
$0.123\pm0.005$, and at the
1994 conferences~\cite{schaile}, \alpmzsq $0.125\pm0.005$.
The change between 1995 and 1996 is due to a combination of
shifts in the values of the \z0 lineshape parameters, redetermined in
light of the recalibration of the LEP beam energy due to the
`TGV effect' \cite{blondel}, and a change in the central value of
$M_{Higgs}$ at which \alpmzsq is quoted, from 300 GeV (1995)
to the fitted value 127 GeV (1997). Studies of
theoretical uncertainties imply~\cite{mor97} that they contribute at
the level of $\pm0.002$ on \alpmzsq.
Since data-taking at the \z0 resonance has now been completed at the
LEP collider the precision of this result is not expected to improve
further.

\begin{figure} [tbh]
   \epsfxsize=4.0in
   \epsfysize=3.0in
   \begin{center}
    \mbox{\epsffile{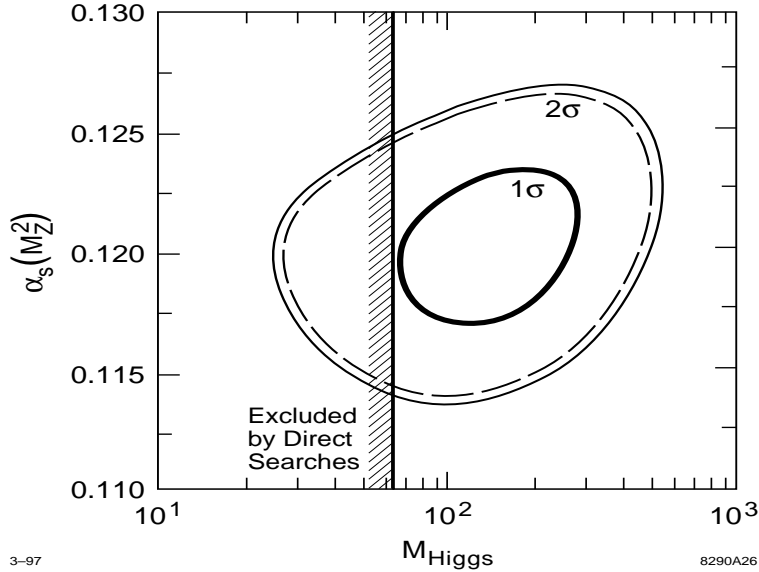}}
\end{center}
   \caption[]{
Results of a global fit of the Standard Model to electroweak 
observables~\cite{blondel};
the 1- and 2-standard deviation contours are shown in the \alpmzsq vs.
$M_{Higgs}$ plane.\hfill$\;$
  }
\end{figure}

\subsection{Hadronic $\tau$ Decays}

An inclusive quantity similar to $R$ is the ratio $R_{\tau}$ of
hadronic to leptonic decay branching ratios, $B_h$ and $B_l$
respectively, of the $\tau$ lepton:
\begin{eqnarray}
R_{\tau}\qu \qu \equiv \qu {B_h \over B_l} \qu
= \qu {1-B_e-B_{\mu} \over B_e}
\end{eqnarray}
where $B_e$ and $B_{\mu}$ can either be
measured directly, or deduced from a measurement of the $\tau$
lifetime $\tau_{\tau}$.
In addition, a family of observables known as `spectral
moments', $R_{\tau}^{kl}$, of the invariant mass-squared
$s$ of the hadronic system has been proposed \cite{spectral}.
 $R_{\tau}$ and $R_{\tau}^{kl}$
have been calculated perturbatively up to \oalpc.
However, because  $M_{\tau}$ $\sim$ 1 GeV one expects (eq.~(1))
\alp($M_{\tau}$) $\sim$ 0.3 and
it is not \apriori obvious that the perturbative calculation can be
expected to be reliable, or that the non-perturbative contributions of 
O($1/M_{\tau}$) will be small.
In recent years a large theoretical effort has been devoted to this
subject; see \eg Refs.~\cite{spectral,bnplp,neubert}.
 
The ALEPH Collaboration derived $R_{\tau}$ from its measurements of
$B_e$, $B_{\mu}$, and $\tau_{\tau}$, and also
measured the (10), (11), (12), and (13) spectral moments.
A combined fit yielded~\cite{alconf} \alpmzsq =
$0.124\pm0.0022\pm0.001$, where the first error receives equal
contributions from experiment and theory, and the second derives from
uncertainties in evolving \alp across the c and b thresholds.
The OPAL Collaboration measured $R_{\tau}$ from
$B_e$, $B_{\mu}$, and $\tau_{\tau}$, and derived \cite{optau}
\alpmzsq=0.1229$^{+0.0016}_{-0.0017}$ (exp.) $^{+0.0025}_{-0.0021}$
(theor.). The CLEO Collaboration measured the same four spectral
moments as ALEPH and also derived $R_{\tau}$ using 1994 Particle
Data Group values for $B_e$, $B_{\mu}$ and $\tau_{\tau}$.
A combined fit yielded \cite{cltau}
\alpmzsq = $0.114\pm0.003$. This central value is slightly lower than
the ALEPH and OPAL values. If more recent
world average values of $B_e$ and $B_{\mu}$ are used CLEO obtains
a higher central \alpmzsq value~\cite{cltau}.
Averaging the second CLEO result and the ALEPH and OPAL
results by weighting with the experimental
errors, assuming they are uncorrelated,
yields:
\begin{eqnarray}
 \alpmzsq\quad =\quad 0.122 \pm 0.001\;\; ({\rm exp.})\;\; \pm0.002\;\; 
({\rm theor.}).
\end{eqnarray}
This is nominally a very precise measurement, although recent studies have 
suggested that additional theoretical uncertainties 
may be as large as $\pm0.006$~\cite{alttau}.
 
\subsection{Hadronic Event Shape Observables}

The rate of 3-jet production, $R_3\equiv\sigma_{3-jet}/\sigma_{had}$, 
is directly proportional to \alp. 
More generally one can define other infra-red- and collinear-safe measures of 
the topology of hadronic final states; for a discussion see 
\eg Ref.~\cite{sldalp}. Such observables are constructed to be
directly proportional to \alp at leading order, and
so are potentially sensitive measures of the strong coupling.
The \oalpsq QCD prediction for each of these observables $X$ can be 
written~\cite{ert}:
\begin{eqnarray}
{1\over\sigma_0} {{\rm d}\sigma\over{\rm d}X}
\quad=\quad A(X)\;\left({\alp\over 2\pi}\right)
\quad+\quad B(X)\left(
{\alp\over 2\pi}\right)^2
\end{eqnarray}
so that \alp can be determined from each. Though these observables are
intrinsically highly correlated, by using many one can
attempt to maximise the use of the information in complicated multi-hadron
events, and in some sense make a more demanding test of QCD than by using
only one or two observables. Moreover, the study of many
observables is essential, as it exposes systematic effects. Finally, the \alp
determination from hadronic event shape observables is based on the 
information content within 3-jet-like events, and is essentially uncorrelated 
with the measurements from the \z0 lineshape which are based on event-counting
of predominantly 2-jet-like final states.

The technology of this approach has been developed
over the past 15 years of analysis at the PETRA, PEP, TRISTAN, SLC and
LEP colliders, so that the method is considered to be well understood
both experimentally and theoretically. Note, however, that before they can be
compared with perturbative QCD predictions,
it is necessary to correct the measured distributions for any
bias effects originating from the detector
acceptance, resolution, and inefficiency, as well
as for the effects of initial-state radiation and
hadronisation, to yield `parton-level' distributions. 

A fit of \oalpsq perturbative QCD to 15 of these observables is shown in 
Fig.~2~\cite{sldalp}. It yields the distressing result
that the \alpmzsq values so determined are not internally consistent with
one another! A measure of the scatter among the results is given by the r.m.s.
deviation of $\pm0.008$, which is much larger than the experimental error of
$\pm0.003$ on a typical observable. One can explain the scatter as an artefact
of the fact that an \oalpsq calculation was employed, and argue that the data
tell us that higher-order terms are needed in order to obtain consistent 
results.
                      
\begin{figure} [tbh]
   \epsfxsize=5.0in
   \epsfysize=3.5in
   \begin{center}
    \mbox{\epsffile{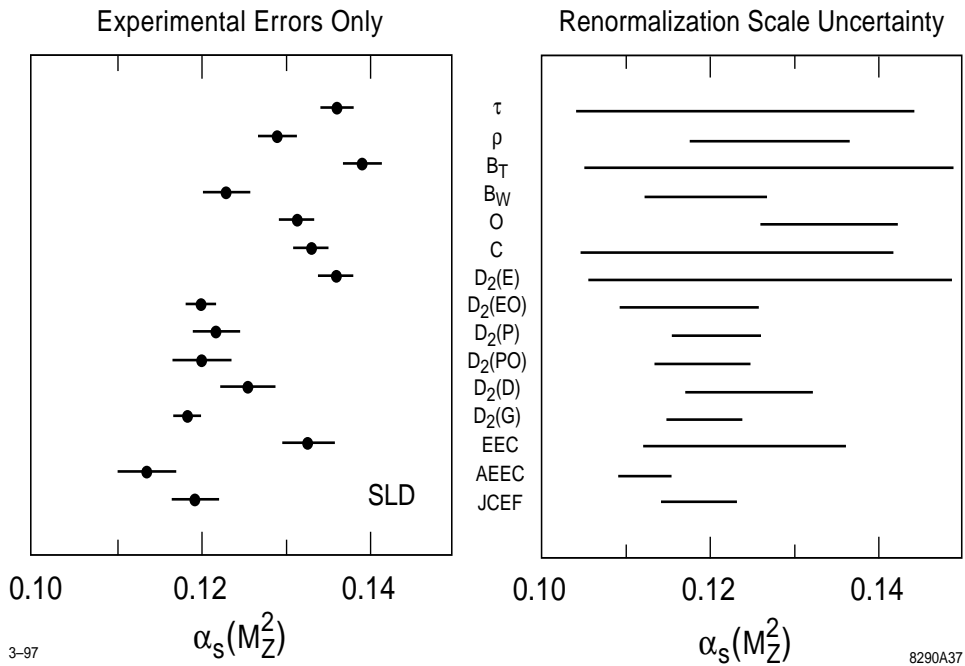}}
\end{center}
   \caption[]{
(a) Values of \alpmzsq determined~\cite{sldalp} 
by fitting \oalpsq QCD predictions to 15 hadronic
event shape observables using a fixed value of the renormalisation scale
$\mu$ = $Q$; the results are clearly inconsistent within the
experimental errors. (b) Renormalisation scale uncertainties.\hfill$\;$
  }
\vskip .5truecm
\end{figure}

A consensus has arisen among experimentalists 
that the effect of such missing higher-order terms can be
estimated from the dependence of \alpmzsq on the value of the renormalisation 
scale $\mu$ assumed in fits of the calculations to the data, and a
renormalisation scale uncertainty is often quoted.
An estimate~\cite{sldalp} of the renormalisation scale uncertainty for each 
observable is shown in Fig.~2b. It is apparent that the scale uncertainty is
much larger than the experimental error, and that the \alpmzsq values are
consistent within these uncertainties.
Though this is comforting, in that it indicates that QCD is self-consistent, 
the necessary addition of large theoretical uncertainties to otherwise precise
experimental measurements is frustrating.
 
For the hadronic
event shape observables \oalpc contributions have not yet been
calculated completely. However, for six observables 
improved calculations can be
formulated that incorporate the resummation \cite{resum}
of leading and next-to-leading
logarithmic terms matched to the \oalpsq results. The matched
calculations are expected {\it a priori} both
to describe the data in a larger
region of phase space than the fixed-order results, and to yield a
reduced dependence of \alp on the renormalization scale, both of  which have 
indeed been found~\cite{sldalp}. 
Application of other approaches to circumvent the scale ambiguity in $\alpha_s$
measurement, involving the use of `optimised' perturbation theory~\cite{opt}
and Pad${\acute{\rm e}}$ Approximants~\cite{pade}, can be found 
in Refs.~\cite{bmmo,bs} respectively.

Hinchliffe has reviewed the various hadronic event shapes-based 
measurements from experiments
performed in the c.m. energy range $10\leq Q\leq 91$ GeV, utilising both
\oalpsq and resummed calculations, and quotes an average value of
\alpmzsq = $0.122\pm0.007$ \cite{ian},
where the large error is dominated by the renormalisation 
scale uncertainty, which far
exceeds the experimental error of about $\pm0.002$.
The LEP experiments have applied similar techniques to determine \alp
from their high-energy data samples at $Q$ = 133, 161 and 172 GeV; results
from OPAL are shown in Fig.~3 \cite{opalhi}. These data points add considerable
lever-arm to tests of the running of \alp.
Schmelling's recent compilation of \alp measurements using event shapes 
includes the 133 GeV results, and
yields~\cite{schmelling} a global average:
$$
\alpmzsq \quad = \quad 0.121 \pm 0.005,
$$
in agreement with Ref.~\cite{ian}, but assuming a more aggressive scale 
uncertainty. 

\begin{figure} [tbh]
   \epsfxsize=4.0in
   \epsfysize=3.0in
   \begin{center}
    \mbox{\epsffile{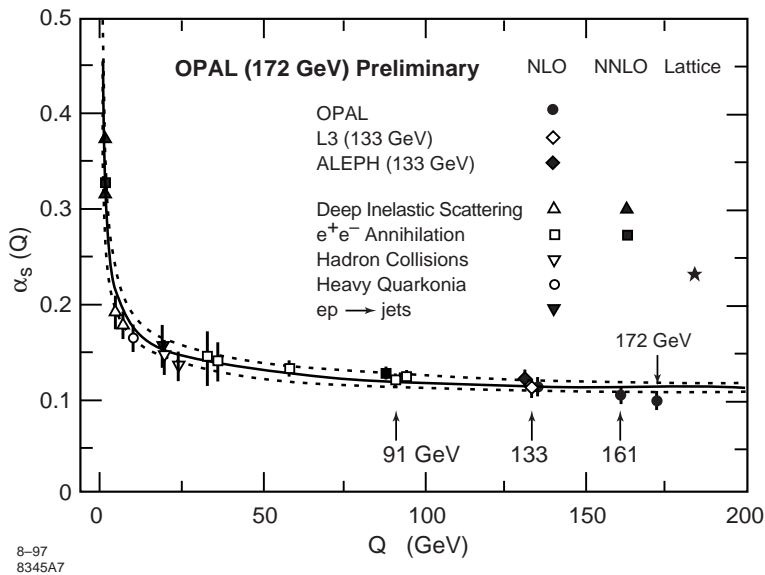}}
\end{center}
   \caption[]{
Measurements of \alp as a function of $Q$, including the latest high-energy
data points~\cite{opalhi}.\hfill$\;$
  }
\end{figure}

Finally, in a recent analysis
the L3 Collaboration has utilised \z0 events with a hard radiated
final-state photon, which reduces the effective c.m. energy available
to the hadronic system, to examine the $Q$-evolution
of four event shape observables in the range $30\leq Q \leq 86$ GeV.
By comparing with resummed + \oalpsq calculations they derived
\cite{l3warsaw} values of \alp in each energy bin (Figure~4), 
which are consistent with \alpmzsq = 0.121.
                                            
\begin{figure} [tbh]
   \epsfxsize=4.0in
   \epsfysize=3.0in
   \begin{center}
    \mbox{\epsffile{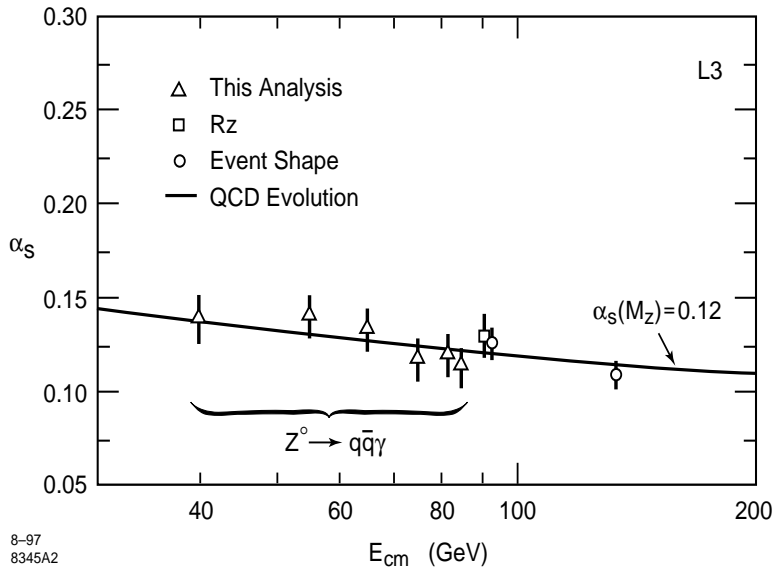}}
\end{center}
   \caption[]{
Measurements of \alp as a function of c.m. energy from L3~\cite{l3warsaw}.
  }
\end{figure}

\subsection{Scaling Violations in Fragmentation Functions}

Though distributions of final-state hadrons are not, in general, calculable
in perturbative QCD, the $Q^2$-evolution of the
scaled momentum ($x_p = p/p_{beam}$) distributions of hadrons, or
`fragmentation functions', can be calculated and used to determine \alp.
In addition to the usual renormalisation scale $\mu$,
a {\it factorisation
scale} $\mu_F$ must be defined that delineates the boundary between the
calculable perturbative, and incalculable non-perturbative, domains.
Additional complications
arise from the changing composition of the underlying event
flavour with $Q$ due to the different $Q$-dependence of the $\gamma$
and \z0 exchange processes. Since B and D hadrons typically
carry a large fraction of the beam momentum, and contribute a large
multiplicity from their decays, it is necessary to consider the scaling
violations separately in b, c, and light quark events, as well as in
gluon jet fragmentation.
 
The ALEPH Collaboration used its \z0 data to constrain
flavour-dependent effects by tagging event
samples enriched in light, c, and b quarks, as well as a sample of
gluon jets~\cite{alfrag}.
The fragmentation functions for the different flavours and the gluon
were parametrised at a reference energy, evolved with $Q$ according to
the perturbative DGLAP formalism calculated at next-to-leading 
order~\cite{dglap}, in conjunction with a parametrisation
proportional to $1/Q$ to represent non-perturbative effects, 
and fitted to
data in the range $22\leq Q\leq 91$ GeV (Fig.~5). They derived
\alpmzsq = $0.126\pm0.007$ (exp.) $\pm0.006$ (theor.), where the
theoretical uncertainty is dominated by variation of the
factorisation scale $\mu_F$
in the range $-1\leq {\rm ln}\mu_F^2/Q^2\leq 1$;
variation of the renormalisation scale in the same range contributed
only $\pm0.002$.
DELPHI has recently reported a similar analysis \cite{newdel} yielding
\alpmzsq = $0.124^{+0.006}_{-0.007}$ (exp.) $\pm0.009$ (theor.).
Combining the ALEPH and later DELPHI results, assuming uncorrelated
experimental errors, yields:
\begin{eqnarray}
\alpmzsq \qu = \qu 0.125 \pm 0.005 ({\rm exp.})\pm0.009({\rm theor.})
\end{eqnarray}
      
\begin{figure} [tbh]
   \epsfxsize=4.0in
   \epsfysize=4.0in
   \begin{center}
    \mbox{\epsffile{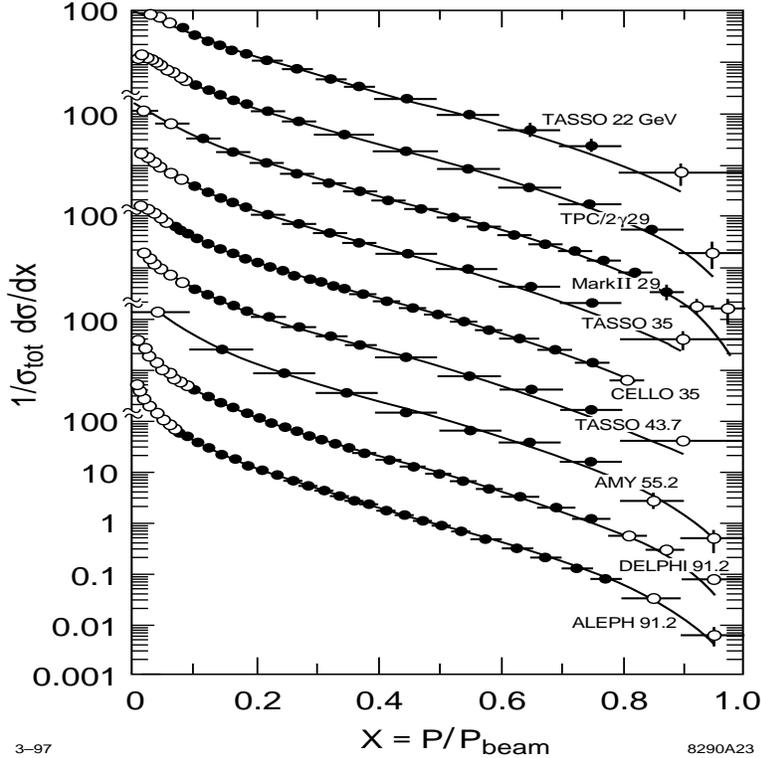}}
\end{center}
   \caption[]{
Inclusive charged-particle fragmentation functions used in \alp 
determination~\cite{alfrag}.\hfill$\;$
  }
\end{figure}
  
\subsection{Comparison with Other Measurements of \alpmzsq}
 
A summary of world \alp measurements, all evolved to $Q$ = $M_Z$, is shown in
Fig.~6~\cite{cracow}. These are drawn from lepton-hadron
scattering, hadron-hadron collisions, heavy quarkonia decays and lattice
gauge theory, as well as \epa. In addition to being relatively precise, 
the \ep results have the invaluable feature
that they bracket the $Q$-range of the experiments, from around 1 GeV 
to 172 GeV, providing the largest
lever-arm for tests of consistency of \alpmzsq measured at different energy
scales. It is clear that, within the uncertainties, all results are consistent
with each another. 
                      
\begin{figure} [tbh]
   \epsfxsize=2.5in
   \epsfysize=1.5in
\includegraphics{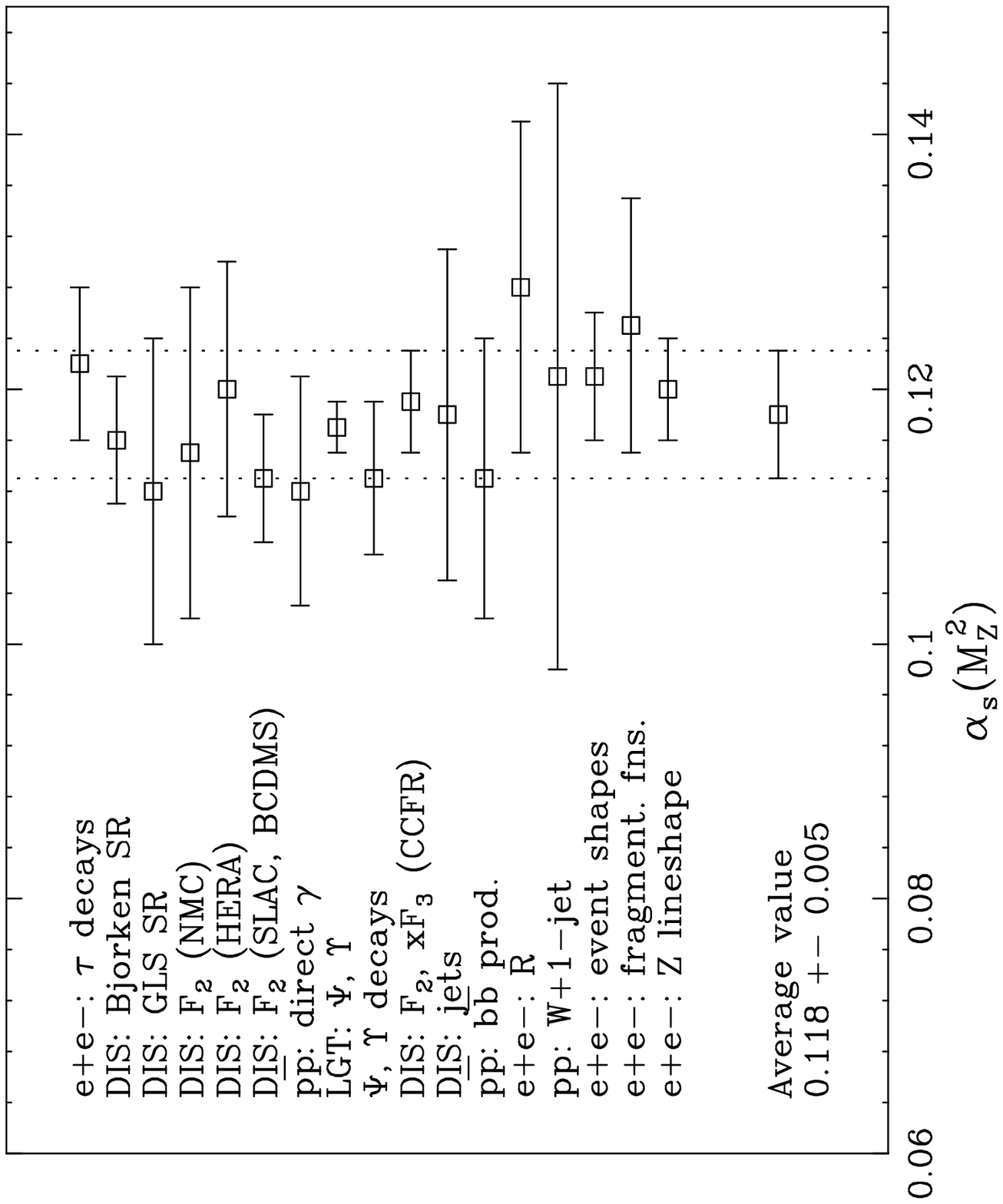}
\vskip 10truecm
   \caption[]{
Summary of world \alpmzsq measurements~\cite{cracow}.
The results are ordered vertically in terms of the hard scale $Q$ of the 
experiment.\hfill$\;$
  }
\end{figure}
 
Taking an average over all 17 measurements
{\it assuming} they are independent, by weighting each by its {\it total}
error, yields \alpmzsq = 0.118 with a $\chi^2$ of 6.4; the low $\chi^2$
value reflects the fact that most of the
measurements are theoretical-systematics-limited. Taking an unweighted average,
which in some sense corresponds to the assumption that all 17
measurements are completely correlated, yields the same result. 
The r.m.s. deviation of the 17
measurements w.r.t. the average value characterises the
dispersion, and is $\pm0.005$. In a quantitative sense, therefore, QCD has
been tested to a level of about 5\%.

\section{3-jet Production}

\subsection{Flavour-Independence of Strong Interactions}

A fundamental assumption of QCD is
that the strong coupling is independent of
quark flavor.  This can be tested by measuring \alp in
events of the type $e^+e^-\rightarrow q{\bar q}(g)$ for specific 
quark flavors $q$. Although an absolute determination of $\alpha_s$
for each quark flavor would have large theoretical
uncertainties, it is possible to test the 
flavor-independence of QCD precisely by measuring ratios 
of couplings in which most experimental errors
and theoretical uncertainties are expected to cancel.  
Since a flavor-dependent anomalous quark chromomagnetic 
moment could modify the probability for the radiation of gluons, 
comparison of the strong	
coupling for different quark flavors may also provide information 
on physics beyond the Standard Model.
 
The development of precise vertex detectors at \ep colliders has allowed
pure samples of b-quark events to be tagged with high efficiency.
Measurements made at LEP of $\alpha^b_s/\alpha_s^{udsc}$                  
have reached precisions between 
$\pm 0.06$ and $\pm 0.013$~ (Fig.~7). However, these tests make the
simplifying assumption that $\alpha_s$ is independent of flavor for all the
non-$b$ quarks, and are insensitive to differences between $\alpha_s$ for
these flavors, especially a different $\alpha_s$ for $c$ quarks compared
with either $b$ or light quarks.  The OPAL Collaboration has measured
$\alpha^f_s/\alpha_s^{all}$  for all five flavors $f$ with no assumption 
on the relative value of $\alpha_s$ for different flavors to
precisions of $\pm 0.026$ for $b$ and $\pm 0.09$ to $\pm 0.20$ for the other
flavors~\cite{opalflav}. 
The kinematic signatures of charmed meson and light hadron 
production that OPAL used to tag $c$- and light-quark events, respectively, 
suffer from low efficiency and strong biases, due to preferential tagging of
events without hard gluon radiation.

The SLD Collaboration has recently presented an update~\cite{sldflav}
of its test of the flavor independence of strong interactions, based on
selection of $Z^0 \rightarrow b \overline{b} (g)$ and $Z^0 \rightarrow q_l 
\bar{q_l}(g)$ ($q_l = u,\, d,\, s$)
events using their quark decay lifetime signatures,
with high efficiency and purity, and with low bias against 3-jet events. 
From a comparison of the 3-jet fractions in each tagged sample with that in
the all-flavours sample, and after unfolding for sample purity and tag bias, 
they obtained~\cite{sldflav}:

$$
\frac{\alpha_s^{uds}} { \alpha_s^{all}} = 
0.997 \pm 0.011\,({\rm stat}) 
          \pm 0.011\,({\rm syst}) 
          \pm 0.005\,({\rm theory})
$$
$$
 \frac{ \alpha_s^c} { \alpha_s^{all}}  =
 0.984 \pm 0.042\,({\rm stat}) 
          \pm 0.053\,({\rm syst}) 
          \pm 0.022\,({\rm theory})
$$
$$
   \frac{ \alpha_s^b} { \alpha_s^{all}} = 
 1.022 \pm 0.019\,({\rm stat}) 
          \pm 0.023\,({\rm syst})
          \pm 0.012\,({\rm theory})  
$$

\noindent
A compilation of the \z0 results is shown in Fig.~7; there is currently no
indication of any flavour dependence of strong interactions.
\vskip.5truecm                      
\begin{figure} [tbh]
   \epsfxsize=4.0in
   \epsfysize=4.0in
   \begin{center}
    \mbox{\epsffile{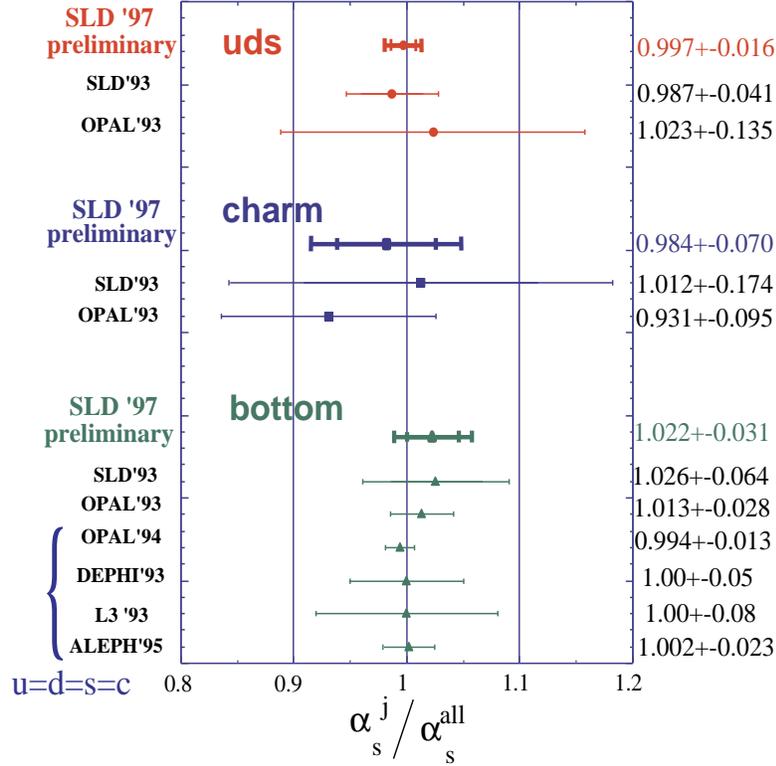}}
\end{center}
   \caption[]{
Summary of \z0 tests of the flavour-independence of strong 
interactions~\cite{sldflav}.\hfill$\;$
  }
\end{figure}

\subsection{Quark- and Gluon-jet Differences}

Many attempts were made at PETRA, PEP and TRISTAN to investigate possible
differences between the properties of quark and gluon jets. These searches
were motivated by the observation that the `colour charge'
of a gluon in a separating octet gg system is 9/4 times that of a quark
in a separating triplet \qq system. It then follows from a leading logarithm
bremsstrahlung-type calculation~\cite{stan} that in the
asymptotic limit $Q$ $\rightarrow$ $\infty$, the multiplicity of soft
gluons in a gluon-initiated jet is 9/4 times the multiplicity in a
quark-initiated jet. Assuming proportionality between the gluon multiplicity
and the ensuing hadron multiplicity leads to the prediction that the
particle multiplicity in gluon jets should be $r$ = 9/4 times that in quark
jets, and hence that the former will have a softer fragmentation function 
by roughly the same
factor. Using similar arguments, it was also shown~\cite{einhorn} that the
angular widths $\delta$ of gluon and quark jets are related by: $\delta_g$ =
$\delta_q^{4/9}$, {\it i.e.}
that gluon jets should be wider than quark jets.
 
However, the experimental searches for these effects, see \eg 
Ref.~\cite{qcd90}, yielded differences in properties significantly smaller 
than the factor of 9/4. It is important to note that there are
several caveats to the naive theoretical predictions which tend to dilute the 
factor $r$. One should consider beyond-leading-order corrections, finite energy
corrections, heavy quark decays, and
fragmentation effects. For jets in \z0 decays, with energy between 30 and
50 GeV, these effects reduce $r$ to the range $1.4<r<1.6$;
see Ref.~\cite{opalqg} and references therein. The
differences in particle multiplicity, width and hardness of fragmentation
are thus expected to be less apparent than the naive prediction.
 
Studies at LEP have established such differences at about the expected 
level~\cite{bill}, but a quantitative comparison with the QCD predictions has
been complicated by the difficulty of relating the experimental jet 
definition and event selection procedures to those assumed in the
calculations. In particular, the calculations assume massless separating 
\qq and gg systems and are completely inclusive, whereas the experimental
studies are based upon the selection of three-jet events using particular
jet-finding algorithms, and the results are algorithm-dependent.

Recently a more consistent analysis procedure has been proposed~\cite{bill2},
and applied by the OPAL Collaboration. The method involves selecting
3-jet final states in which two heavy-quark jets recoil in the same hemisphere
against the third (gluon) jet. After correcting for misidentification, the
properties of jets in this gluon sample were then compared with those of a 
sample of back-to-back two-jet light-quark \qq events, tagged on the basis of
the absence of long-lived (heavy-quark) decay products. Because of the
kinematic bias caused by the 3-jet topology the 278 gluon-tagged jets had a
mean energy of 39.2 GeV, compared with 45.6 GeV for the sample of roughly
28,000 light-quark jets. This small difference was corrected assuming the QCD
energy-dependence of the mean multiplicity, and yielded~\cite{opalqg} 
$r(39 {\rm GeV})$ = $1.552\pm0.041$ (stat) $\pm0.060$ (syst.). This result is
shown in Fig.~8~\cite{opalqg}, where it is compared with the analytic QCD 
calculations. The measurement is in good agreement with the 
next-to-next-to-leading order calculation that includes energy/momentum
conservation. This important result helps to resolve the long-standing
confusion over the low measured values of $r$, and gives us further confidence
that QCD is able to describe the inclusive properties of hadronic jets.

\begin{figure} [tbh]
   \epsfxsize=4.0in
   \epsfysize=3.0in
   \begin{center}
    \mbox{\epsffile{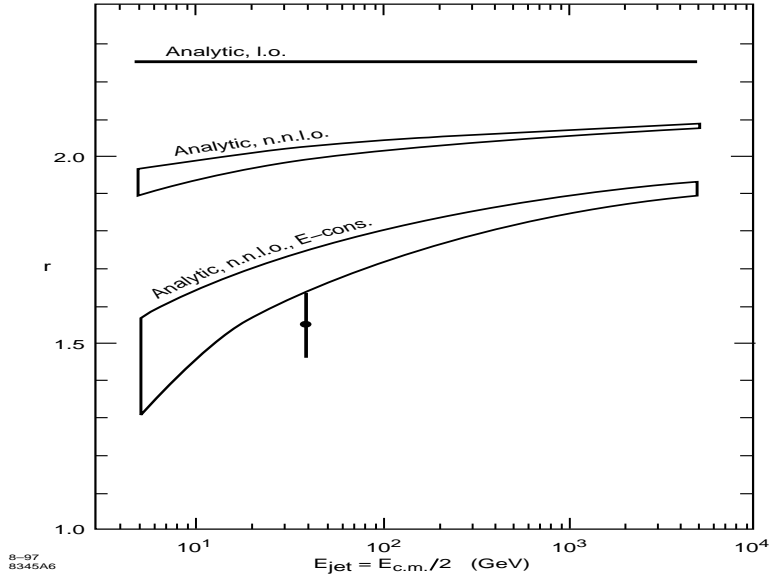}}
\end{center}
   \caption[]{
The ratio of particle multiplicity in gluon and quark jets vs. jet 
energy~\cite{opalqg}.\hfill$\;$
  }
\end{figure}

\subsection{Search for T$_N$-odd Effects in \z0 \ra q$\overline{\rm q}$g}

The $Z^0$ bosons produced at SLC using
longitudinally polarized electrons have polarization
along the beam direction $A_Z = (P_{e^-}-A_e)/(1-P_{e^-}\cdot A_e)$,
where  $P_{e^-}$ is
the electron beam polarization, defined to be negative (positive) for a
 left-(right-) handed
beam, and $A_e = 2v_ea_e/(v_e^2+a_e^2)$ with $v_e$ and $a_e$ the
electroweak vector and axial
vector coupling parameters of the electron, respectively. 
An electron-beam polarization at the \ep~interaction point
of approximately 0.77 in magnitude was achieved
in the 1994-95 and 1996 runs, yielding $A_Z$ = $-0.82$ ($+0.71$) for $P_{e^-}$ =
 $-0.77$ ($+0.77$) respectively. 
For polarized $Z^0$ decays to
three hadronic jets one can define the triple-product:
$\vec{S_Z}\cdot(\vec{k_1}\times \vec{k_2})$,
which correlates the \z0 boson polarization vector $\vec{S_Z}$
with the normal to the three-jet plane defined by
$\vec{k_1}$ and $\vec{k_2}$, the momenta of the highest- and the
second-highest-energy jets respectively.
The triple-product is even under $C$ and $P$ reversals, and odd
under $T_N$, where $T_N$ reverses momenta and spin-vectors without
exchanging initial
and final states. Since $T_N$ is not a true time-reversal operation a
non-zero value does not signal $CPT$ violation and is possible in a theory that
respects $CPT$ invariance.

The tree-level
differential cross section for $e^+e^- \rightarrow q{\bar q}g$ for a
longitudinally polarized
electron beam and massless quarks may be written \cite{Brandenburg}:
\begin{equation}
{1 \over \sigma}
{{d\sigma} \over {d \cos\omega}} = {{9} \over {16}} [ (1-{1\over
 3}\cos^2\omega)+\beta
\, A_Z \, \cos\omega],
\label{diff}
\end{equation}
where $\omega$ is the polar angle of the vector normal to the jet plane,
$\vec{k_1}\times \vec{k_2}$, w.r.t. the electron beam direction.
With $\beta |A_Z|$ representing the magnitude, 
the second term is proportional to the $T_N$-odd triple-product, 
and appears as a forward-backward asymmetry of the
jet-plane-normal relative to the \z0 polarization axis.
The sign and magnitude of this term are different for the two beam helicities.
Recently Standard Model $T_N$-odd contributions of this form
at the \z0 resonance have investigated~\cite{Brandenburg}. The triple-product
vanishes identically at tree level, but non-zero contributions arise
from higher order processes. Due to various cancellations these contributions
are found to be very small at the $Z^0$ resonance and yield 
$ |\beta | \sim 10^{-5}$.
Because of this background-free situation, measurement of the
cross section (\ref{diff}) is
sensitive to physics processes beyond the Standard Model that give
$\beta \neq$ 0.
 
The SLD measurement of the $\omega$ distribution in all-flavours \z0 decays 
is shown in Fig.~9~\cite{sldtodd}. 
A fit of Eq.~(8) yields $\beta=0.008\pm0.015$, or $-0.022<\beta<0.039$ at 95\%
c.l. Similar measurements for the potentially more interesting \bbg system are 
now in progress~\cite{sldtoddb}.
 
\begin{figure} [tbh]
   \epsfxsize=4in
   \epsfysize=3in
   \begin{center}
    \mbox{\epsffile{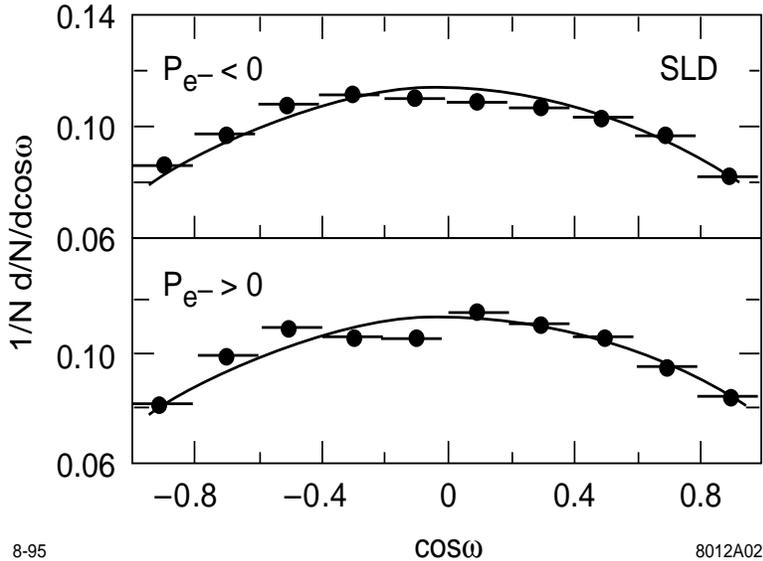}}
\end{center}
   \caption[]{
SLD measured distributions of cos$\omega$; the line is a fit of Eq.~(8).
\hfill$\;$
  }
\end{figure}

\section{4-jet Production}

\subsection{Total Cross-Section}

It has been known since studies performed at PETRA at c.m. energies of
approximately 35 GeV, see \eg Ref.~\cite{tasso4}, that the tree-level \oalpsq
matrix element calculation of the 4-jet cross-section is insufficient to
describe the data, Fig.~10. Tremendous progress has been made recently and the
full 1-loop calculation has been performed~\cite{dixon4}. A comparison of the
leading-$N_C$ contributions to the 4-jet rate measured with \z0 data and a
similar jet algorithm as in Fig.~10 is given in Fig.~11. The 1-loop result
is roughly a factor of two larger than the tree-level result and
describes the data well; sub-leading
$N_C$ contributions are typically an order of magnitude smaller~\cite{dixon4}.
 
\begin{figure} [tbh]
   \epsfxsize=4in
   \epsfysize=3in
   \begin{center}
    \mbox{\epsffile{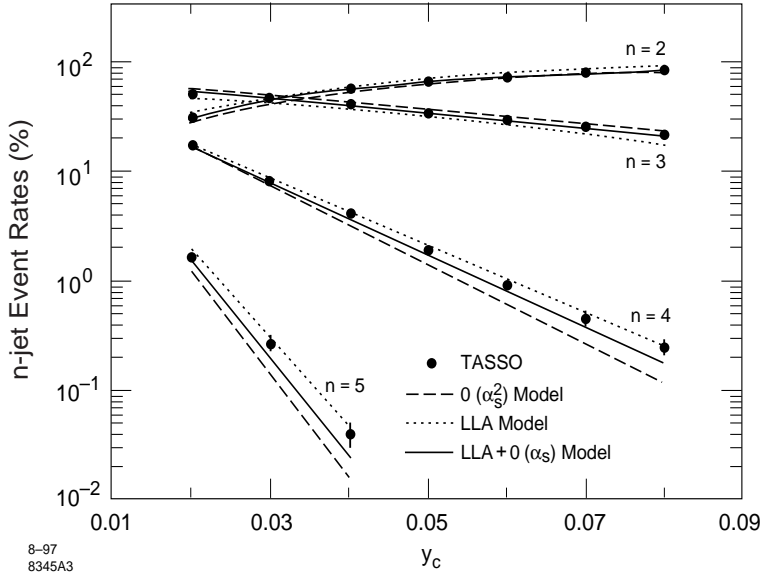}}
\end{center}
   \caption[]{
$n$-jet event rates measured at $Q$ = 35 GeV compared with QCD calculations;
the \oalpsq calculation is unable to describe the 4-jet 
cross-section~\cite{tasso4}.
\hfill$\;$
  }
\end{figure}
 
\begin{figure} [tbh]
   \epsfxsize=4in
   \epsfysize=3in
   \begin{center}
    \mbox{\epsffile{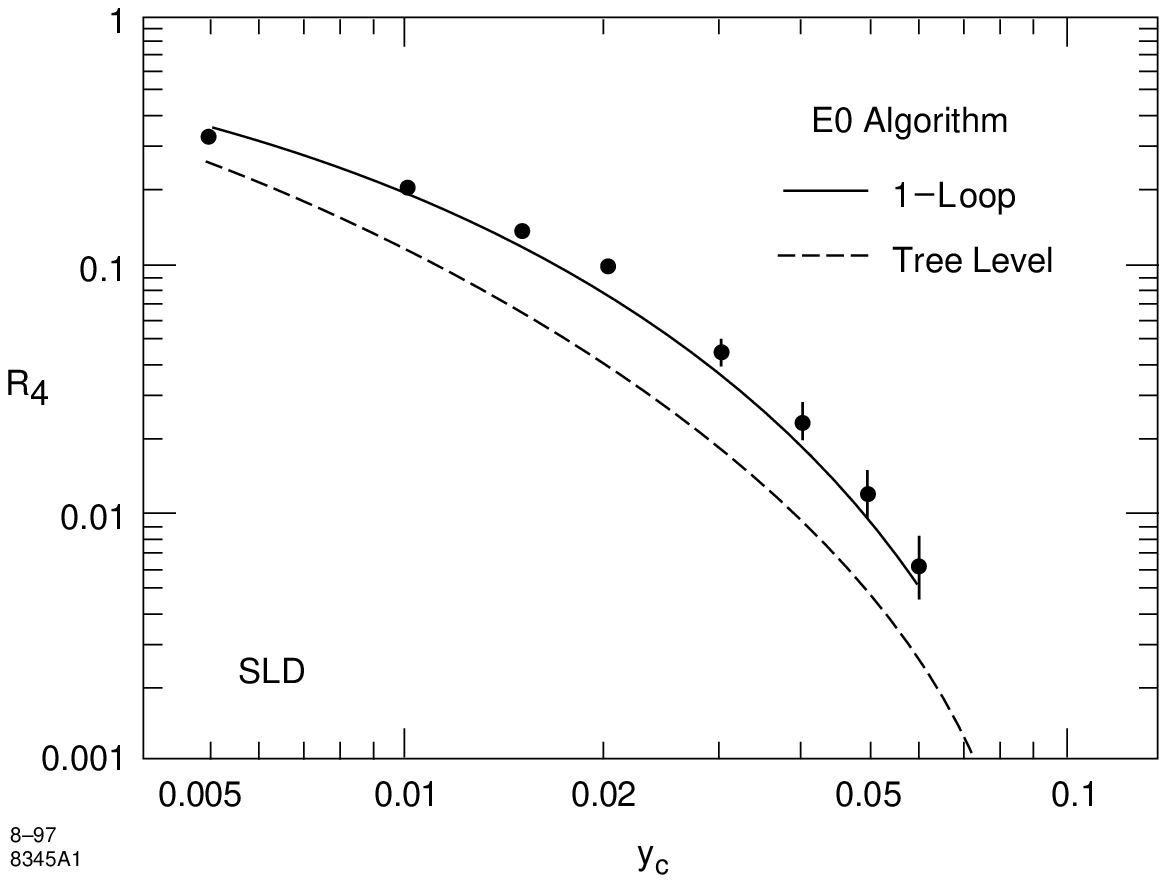}}
\end{center}
   \caption[]{
$4$-jet event rate measured at $Q$ = 91 GeV compared with QCD calculations;
the \oalpc calculation describes the data~\cite{dixon4}.\hfill$\;$
  }
\end{figure}

\subsection{Angular Correlations}

The QCD tree-level couplings contributing to 4-jet events are shown in Fig.~12;
they may be classified in terms of the Casimir factors $C_F$, $T_F$, and $N_C$
that characterise the SU(3)$_C$ group; see \eg Ref.~\cite{ssi}.
It is interesting to consider whether the Casimir factors can be measured.
Clearly nature does not deliver events corresponding to the
tree-level vertices shown in Fig.~12! Instead, one must write down the 
Feynman amplitudes for the 4-jet event diagrams, add them to
those for 2- and 3-jet production at the same order of perturbation theory, 
and square them to derive the total hadronic cross section. 
The terms corresponding to 4-jet production can then be identified in a 
gauge-invariant manner, and yield a differential cross section of the form:
$$
{1\over \sigma_0} d\sigma^4\qu = \qu \left({\alp C_F \over \pi}\right)^2
\left[ F_A + \left(1-{1\over 2}{N_C\over C_F}\right) F_B +
{N_C \over C_F} F_C
\right]
$$
\begin{eqnarray}
\qu\qu\qu + \qu \left({\alp C_F \over \pi}\right)^2
\left[{T_F\over C_F} N_f F_D +
\left(1-{1\over 2}{N_C\over C_F}\right) F_E \right]
\end{eqnarray}
where $F_A$ $\ldots$ $F_E$ are kinematic factors. The overall
normalisation of the cross-section is proportional to $(\alp\,C_F)^2$,
and the kinematical distribution of the four jets depends on the ratios
$N_C/C_F$ and $T_F/C_F$, which can hence in principle be measured.

A number of 4-jet angular correlation observables that are potentially 
sensitive to these ratios 
have been proposed~\cite{angcorr}. If one orders and labels 
the four jets in an event in terms of their momenta (or energies) such that
$p_1>p_2>p_3>p_4$ one can define the Bengtsson-Zerwas angle:
\begin{eqnarray}
\cos\chi_{BZ}\quad\propto\quad(\vec{p_1}\times \vec{p_2})\cdot
(\vec{p_3}\times \vec{p_4})
\end{eqnarray}
and the Nachtmann-Reiter angle:
\begin{eqnarray}
\cos\theta^*_{NR}\quad\propto\quad(\vec{p_1}- \vec{p_2})\cdot
(\vec{p_3}- \vec{p_4}).
\end{eqnarray}
Interestingly the 1-loop corrections discussed in the previous section do not
change the shapes of the predicted distributions of these angles~\cite{dixon4},
so that
use of the tree-level calculations, which has been done to date, would appear 
to be valid.
 
 
\begin{figure} [tbh]
\begin{minipage}[t]{2.8in}
   \epsfysize=3.0in
   \begin{center}
   \mbox{\epsffile{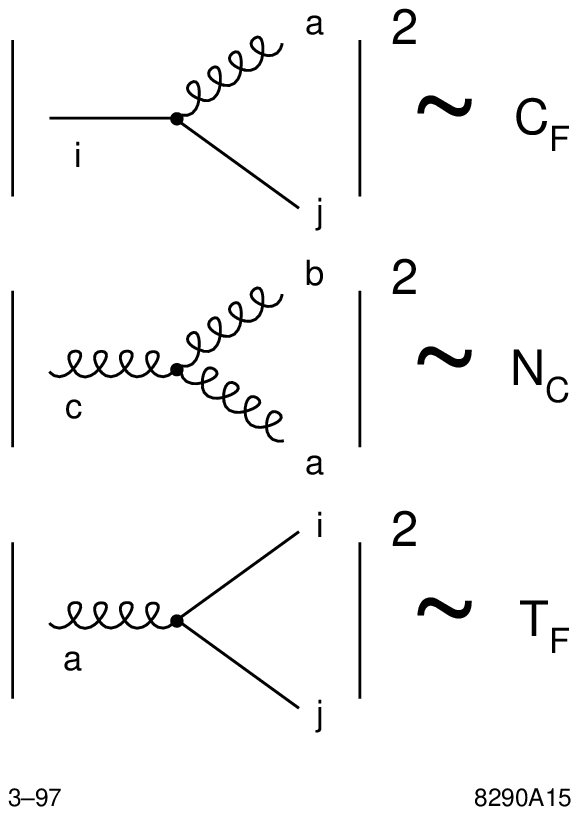}}
   \caption[]{
Casimir classification of tree-level QCD couplings.\hfill$\;$
  }
\end{center}
\end{minipage}
\begin{minipage}[t]{2.8in}
   \epsfysize=2.5in
   \begin{center}
    \mbox{\epsffile{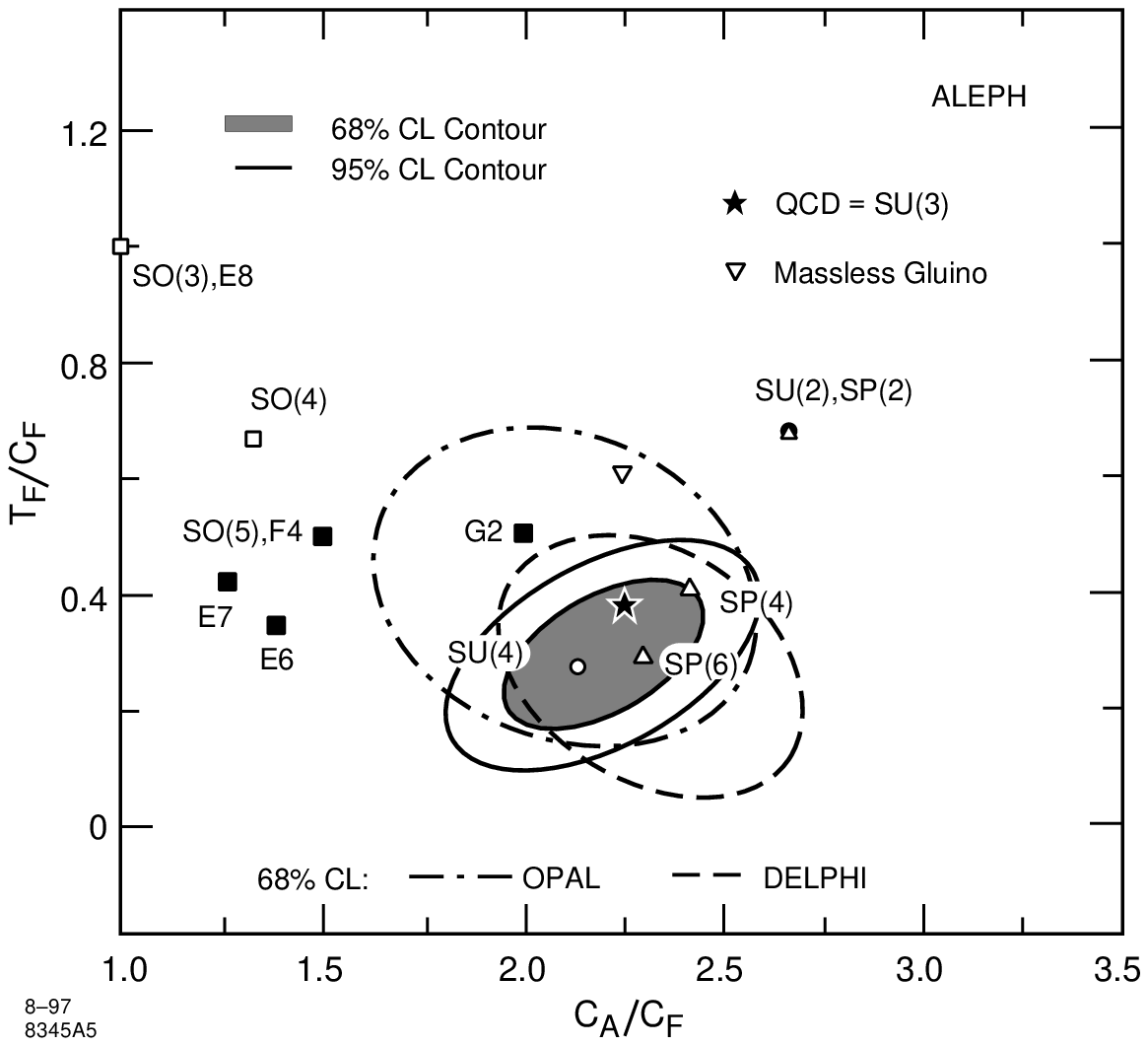}}
   \caption[]{
The $T_F/C_F$ vs. $N_C/C_F$ plane showing recent
measurements~\cite{alephgluino} from LEP
experiments, as well as the expectations from numerous gauge 
groups.\hfill$\;$
  }
\end{center}
\end{minipage}
\end{figure}

In studies by the LEP experiments, fits for $N_C/C_F$ and $T_F/C_F$ have been 
performed to these angular
distributions, as well as to the angle $\alpha_{34}$ between jets 3 and 4.
Results from the most recent study, by the ALEPH Collaboration, 
are displayed in Fig.~13~\cite{alephgluino},
where they are compared with the values for numerous gauge groups.
The SU(3) QCD expectation is clearly in good agreement with the data.
The expectations from several other gauge models, such as SU(4), SP(4) and
SP(6), also appear to be compatible with the experimental results.
Note, however, that none of these models contains three colour degrees of
freedom for quarks, and hence all can be ruled out on that basis. Besides
SU(3), only the U(1)$_3$ and SO(3) models contain three quark colours, but
both are inconsistent with the measured values of $N_C/C_F$ and $T_F/C_F$.
The results shown in Fig.~13 hence yield the remarkable conclusion that SU(3)
is the only known viable gauge model for strong interactions.
Also shown in this figure is the theoretical expectation for QCD augmented with
a single family of light gluinos~\cite{gluino}; the ALEPH result appears to 
rule out
this expectation at better than 95\% confidence level, but this interpretation
has been criticised~\cite{glennys}.


\section{Conclusions}

Electron-positron annihilation is an ideal laboratory for
strong-interaction measurements. Jets in the final state allow the dynamics 
of quarks and gluons to be measured precisely. Since the PETRA/PEP era the
\z0 experiments have established the gauge structure of strong interactions via 
measurement of the Casimir factor ratios $N_C/C_F$ and $T_F/C_F$, 
leading us to the conclusion that QCD {\it is} the correct theory. 
Differences between quark- and gluon-jets of the same energy have been 
convincingly demonstrated and, when compared in a consistent fashion, have been
found to be in agreement with theoretical expectations. 
The coupling \alp has been determined to the 5\%-level of accuracy
from inclusive \z0 lineshape observables
and from hadronic $\tau$ decays, as well as from 
event shape measures and scaling violations in inclusive single-particle 
fragmentation functions. These \alpmzsq measurements are internally consistent,
and agree with results from lepton-nucleon scattering, hadron-hadron
collisions, and lattice gauge theory determined across a wide range of energy
scales.

The development of precise silicon vertex detectors has allowed 
flavour-dependent properties to be studied in both the primary hard process and
in jet fragmentation, and the strong coupling has been found to be 
flavour-independent at the sub-5\% level. In addition, high electron-beam 
polarisation at
SLC has allowed interesting new symmetry tests using 3-jet events. Finally,
tremendous theoretical effort has resulted in perturbative QCD calculations 
that are accurate at the 10\%-level, and attempts to calculate 
non-perturbative effects for jet observables are well under way.
   
\section*{Acknowledgments}
I thank G.~Cowan, W.~Gary, D.~Muller and A. Mnich for useful discussions, and my
colleagues in the SLD Collaboration for their support.

\end{document}